\documentclass[preprint,pre,nofootinbib,showpacs,titlepage]{revtex4}%
\usepackage{amsfonts}
\usepackage{amsmath}
\usepackage{amssymb}
\usepackage{graphicx}
\usepackage{acronym}%
\begin{document}
\title{Pattern formation by boundary forcing in convectively unstable, oscillatory
media with and without differential transport}
\author{Patrick N. McGraw and Michael Menzinger}
\affiliation{Department of Chemisty, University of Toronto, Toronto ON, M5S 3H6, \ Canada}

\begin{abstract}
Motivated by recent experiments and models of biological segmentation, we
analyze the excitation of pattern-forming instabilities in convectively
unstable reaction-diffusion-advection systems, occuring by constant or
periodic forcing at the upstream boundary. Such boundary-controlled pattern
selection is a generalization of the flow-distributed-oscillation (FDO)
mechanism that can be modified to include differential diffusion (Turing) and
differential flow (DIFI) modes. Our goal is to clarify the relationships among
these mechanisms in the general case where there is differential flow as well
as differential diffusion. We do so by analyzing the dispersion relation for
linear perturbations and showing how its solutions are affected by
differential transport. We find a close relationship between DIFI and FDO
modes, while the Turing mechanism gives rise to a distinct set of unstable
modes. Finally, we illustrate the relevance of the dispersion relations using
nonlinear simulations and we discuss experimental implications of our results. \ 

\end{abstract}
\maketitle

\section{Introduction}

Recently, \ theoretical \cite{Andresen}-\cite{Vasquez} and
experimental\ \cite{Kaern1},\cite{Bamforth3}-\cite{Santiago} attention has
been focused on spatiotemporal instabilities in one-dimensional reactive
flows. \ Among these pattern-forming instabilities are the differential flow
(DIFI)\cite{Rovinsky92}\cite{Rovinsky93}\cite{MRDIFI}\cite{Wu},
Turing\cite{Turing}\cite{Kapral}, and the physically distinct flow-distributed
oscillation (FDO)\cite{Andresen}\cite{Kaern1}\cite{Kaern4}\cite{Kaern5}%
\cite{Faraday} mechanisms. Two of these, DIFI and FDO, necessarily involve a
flow, while Turing and DIFI necessarily involve the differential transport of
activator and inhibitor species. \ 

Instabilities in a flowing medium may be absolute or
convective.\cite{Andresen}\cite{Bamforth1}\cite{Proctor}\cite{Deissler}. \ In
the first case, a localized disturbance grows with time and spreads both
upstream and downstream. \ In the convective case, on the other hand, \ a
localized disturbance can not propagate upstream, and so the effect of a
temporary localized perturbation is eventually washed downstream and out of
the system if there is a downstream boundary. \ \ However, persistent
disturbances upstream can have a large effect on the downstream behavior.
\ \ This leads to the possibility of noise-sustained structures\cite{Proctor}%
\cite{Deissler} or patterns which are controlled primarily by the
\emph{upstream} boundary conditions. We are insterested here in this latter
case, where the upstream boundary is crucial to the control of the pattern. \
FDO is a convective mechanism of pattern formation whereby an open flow maps
the temporal dynamics of an oscillating medium, whose phase is set at the
upstream boundary, onto space. \ In the limit of vanishing diffusion, the
resulting stationary\cite{Andresen}\cite{Bamforth1}\cite{Bamforth2}%
\cite{Kaern1}\cite{Kaern5}, travelling\cite{Kaern1}\cite{Kaern2}\cite{Kaern5}
and pulsating\cite{Kaern2}\cite{Kaern5} waves are simple kinematic phase
waves\cite{Kaern1}, making FDO conceptually the simplest of the
pattern-forming mechanisms, although it was discovered later than the others.
\ The Turing instability, by contrast, was initially conceived of as an
absolute instability of a stationary reaction-diffusion medium. \ In a flow
system, however, Turing \cite{Faraday} and DIFI \cite{RazvanNew} patterns can
also be generated and controlled by means of the upstream boundary condition
under convectively unstable conditions.\ 

Since an open flow with a fixed upstream boundary is equivalent, via a
Galilean transformation, to a stationary medium with a moving
boundary\cite{Kaern3}\cite{Kaern4}\cite{Faraday}\cite{Kaern6}, the physical
ideas of FDO and other boundary-driven convective instabilities are also
applicable to growing media. \ In developmental biology an FDO mechanism
driven by an oscillator or "segmentation clock" \ at the growing tip of an
embryo leads to the formation of somites \cite{Wolpert}, \ the precursors of
vertebrae and body segments during early embryogenesis\cite{Kaern3}%
\cite{Kaern4}\cite{Faraday}\cite{Kaern6} (the best-studied examples are chick
and mouse.). Quite generally, the issue of pattern formation on a growing
domain is vitally important to developmental biology.\cite{Murray} \ Recent
laboratory experiments \cite{Kaern6}\cite{Santiago}\cite{Miguez1} in Turing or
Hopf unstable media also make use of a moving boundary that mimics a flow.
\ By contrast, a packed bed reactor (PBR) is a flow reactor in which the
inlet, not the medium, is fixed in the laboratory frame of reference. \ In the
experiments of \cite{Kaern1}\cite{Bamforth3}\cite{Toth}\cite{Kaern2}%
\cite{Kaern5} the reactor is fed by the outlet of a continuous stirred tank
reactor (CSTR) which can be made to oscillate, generating travelling waves in
the PBR-tube, or remain at a fixed point, leading to stationary waves. \ 

An extensive comparison of the parameter ranges for the production of
stationary waves by means of various instabilities was made by Satnoianu et
al.\cite{Satnoianu2}, who suggested that all of these waves be viewed as
variants of a general mechanism called "flow and diffusion- distributed
structures" \ (FDS). \ \ In reference \cite{RazvanNew}, \ travelling waves and
combinations of differential flow and diffusion were also considered.
\ Travelling waves were refered to as DIFI\ waves while stationary waves were
referred to as FDS. \ 

Our goal is to clarify the relationships among the convectively driven FDO and
differential transport (Turing and DIFI) modes in an open flow. We develop a
general linear stability analysis for convective modes driven by boundary
perturbations. \ We illustrate the relationships visually by plotting
solutions of the dispersion \ relations. \ Our approach differs from that of
\cite{Satnoianu2} and \cite{RazvanNew} in several ways. \ First, we choose to
focus on patterns driven convectively by the upstream boundary condition,
distinguishing them from absolute instabilities. We do this because the
possibility of convective instability embodies much of the new behavior that
is possible with a flow (or growth) as opposed to a stationary medium.
\ Accordingly, we treat the dispersion relation for small disturbances
differently, \ taking the real frequency, set by the boundary condition, as
the independent variable and examining the \emph{spatial} behavior of the
resulting disturbance rather than examining the temporal behavior of an
imposed spatial perturbation. \ We consider a mode unstable if it grows with
downstream \emph{distance} in response to a constant or periodic driving at
the boundary.\ \ This approach resembles that of \cite{Andresen} and
\cite{McGraw}. \ 

Within this approach we find it useful to distinguish wave modes not by
whether they are stationary or travelling as in \cite{RazvanNew} but by other
criteria including the phase velocity and the relative phase between
oscillations of the activator and inhibitor. \ We find that the FDO and
DIFI\ mechanisms are closely related to each other, both being related to an
underlying Hopf instability, while the Turing mechanism gives rise to a
distinct set of modes. \ The two sets of modes are apparent as two distinct
peaks in the spatial growth rate at different perturbation frequencies.
\ \ The DIFI/FDO modes can be either travelling or stationary, while Turing
modes are stationary only in the case of zero flow velocity. \ \ In the
zero-flow case (the original case in which the Turing mechanism was
considered), the instability is absolute and therefore not controlled by the
boundary. \ However, it has been observed that Turing patterns can be
generated in a system with nonzero flow \cite{Faraday}\cite{Kaern6}%
\cite{Santiago} in which case they are advected along with the flow, i.e.,
stationary in the co-moving frame. \ In this case the instability can be
convective and a Turing mode with a particular wavelength can be selected by
imposing a periodic perturbation at the inflow. \ We find that in the presence
of simultaneous differential flow and differential diffusion (relevant to the
packed-bed reactor) \ some of the distinguishing features of Turing modes are
modified, but the essential picture of two separate peaks remains unchanged. \ \ 

At the end of the paper, we describe some nonlinear simulations which help to
illustrate some of the linear results and relationships we describe. \ We find
that the linear results give a rather good insight into the nature of the
fully nonlinear solutions, at least in the case where the nonlinearity is not
very strong. \ 

\section{Linear analysis of RDA equations}

We consider the reaction-diffusion-advection \ (RDA) equations describing the
transport and chemical kinetics of an activator and an inhibitor species.
\ The chemical medium is defined by the \textquotedblleft
local\textquotedblright\ or batch reactor dynamics together with transport
terms. We wish to consider several forms of differential transport, so we
allow each species to have its own flow velocity and diffusion coefficient.
\ The RDA equations are: \ \
\begin{align}
\frac{\partial A}{\partial t}  &  =f(A,B)-\phi_{A}\frac{\partial A}{\partial
x}+D_{A}\frac{\partial^{2}A}{\partial x^{2}}\label{RDAeq}\\
\frac{\partial B}{\partial t}  &  =g(A,B)-\phi_{B}\frac{\partial B}{\partial
x}+D_{B}\frac{\partial^{2}B}{\partial x^{2}}.\nonumber
\end{align}
Our aim is to analyze the pattern-forming instabilities, so we shall assume
that the local kinetics has a stable or unstable fixed point, and linearize
the equations for small perturbations of the uniform fixed point state. \ For
convenience, we shall use units in which the flow velocity of species B is
unity. \ Linearizing near the fixed point $(A_{0},B_{0})$, transforming the
units to ones where $\phi_{B}=1,$ and defining the velocity and diffusion
ratios $\delta_{v}=\phi_{A}/\phi_{B}$ and $\delta_{D}=D_{A}/D_{B}$
respectively, and $D=D_{B}$ gives: \ \
\begin{align}
\frac{\partial a}{\partial t}  &  =-\delta_{v}\frac{\partial a}{\partial
x}+\delta_{D}D\frac{\partial^{2}a}{\partial x^{2}}+a_{11}a+a_{12}%
b\label{RDALin}\\
\frac{\partial b}{\partial t}  &  =-\frac{\partial b}{\partial x}%
+D\frac{\partial^{2}b}{\partial x^{2}}+a_{21}a+a_{22}b.\nonumber
\end{align}
where the matrix
\[
\frac{\partial(f,g)}{\partial(a,b)}=%
\begin{pmatrix}
a_{11} & a_{12}\\
a_{21} & a_{22}%
\end{pmatrix}
\]
is the Jacobian of the local kinetic system evaluated at the fixed point and
$a=A-A_{0}$, $b=B=B_{0}$ are the perturbations. \ A complex exponential
solution%
\begin{equation}%
\begin{pmatrix}
a\\
b
\end{pmatrix}
=%
\begin{pmatrix}
u\\
v
\end{pmatrix}
e^{iwt+kx} \label{complexexp}%
\end{equation}
represents a travelling wave in which the concentrations of both species
oscillate.\footnote{The phase convention of the wavenumber $k$ is that of
\cite{Satnoianu2}, \ chosen for later convenience. \ $\operatorname{Re}k$
represents the spatial growth rate, while -$\operatorname{Im}k$ is the inverse
wavelength or \textquotedblleft real\textquotedblright\ wavenumber. \ } \ The
relative amplitude and phase are determined by the complex amplitudes $u$ and
$v.$ \ (A real solution is formed from (\ref{complexexp}) and its complex
conjugate.)\ Substitution into (\ref{RDALin}) gives%
\begin{align}
i\omega u  &  =-\delta_{v}ku+\delta_{D}Dk^{2}u+a_{11}u+a_{12}%
v\label{doubledisp}\\
i\omega v  &  =-kv+Dk^{2}v+a_{21}u+a_{22}v.\nonumber
\end{align}
which can be combined to give the dispersion relation%
\begin{align}
0  &  =\delta_{D}D^{2}k^{4}-(\delta_{D}+\delta_{V})Dk^{3}+D\left(
a_{11}+\delta_{D}a_{22}+\delta_{V}/D-i\omega(\delta_{D}+\delta_{V})\right)
k^{2}+\label{dispersion}\\
&  \left(  -(a_{11}+\delta_{V}a_{22})+i\omega(1+\delta_{V})\right)
k+\Delta-i\omega Tr-\omega^{2},\nonumber
\end{align}
where $Tr=a_{11}+a_{22}$ and $\Delta=a_{11}a_{22}-a_{12}a_{21}$ are
respectively the trace and determinant of the Jacobian. \ \ Two particular
cases of differential transport were studied in previous work. \ The case
$\delta_{V}=1,$ $\delta_{D}\neq1$ is relevant to \cite{Santiago}, in which
there is differential diffusion due to the immobilization of one of the
species, but the flow velocities are the same, since it is actually the
boundary that moves relative to the medium. \ On the other hand, Satnoianu et
al.\cite{Satnoianu2} \ considered the case $\delta_{V}=\delta_{D}\neq1,$
\ which may be a good approximation in a flow system when one of the species
is immobilized and the other moves freely. The pure FDO case $\delta
_{V}=\delta_{D}=1$ has no differential transport. With the appropriate changes
of variables and restrictions on the transport ratios, \ the above dispersion
relation reduces to the ones given in the previous references \cite{Andresen}%
\cite{Faraday}\cite{Satnoianu2}\cite{RazvanNew}\cite{McGraw} for particular
cases. \ \ We wish to consider more general forms of differential transport,
for both theoretical and experimental reasons. \ \ First, varying the two
transport ratios independently allows a fuller understanding of the effects of
the two types of differential transport and their interaction. \ Second, we
wish to allow the possibility of experiments in which the transport
coefficients are related in ways other than those previously considered. \ 

We wish to analyze the steady-state \ response of the system to a sinusoidal
forcing of the inflow boundary at a constant amplitude. \ In general, in the
linear approximation, this gives rise to a travelling wave with a complex wave
number. \ The frequency $\omega$ will be taken to be purely real, reflecting
the constant amplitude of the forcing. \ \ However, the convective dynamics of
the medium may cause the disturbance to grow or damp with the downstream
distance, so that $k$ may be complex. \ Thus, we consider the real $\omega$ as
an independent variable and solve the dispersion relation numerically for the
complex $k$. \ The dispersion relation is quartic in $k$ and so has in general
four solutions. \ Each is associated with an eigenvector $\mathbf{u=}%
\begin{pmatrix}
u\\
v
\end{pmatrix}
$\ \ which can be found by substituting the solution $k$ back into
(\ref{doubledisp}). \ In this way we can find $k(\omega)$ and
$\mathbf{u(\omega).}$ \ \ The ratio $R(\omega)\equiv v/u$, \ which in general
is complex, gives information about the relative amplitude and phase of
oscillations in the two species (an example is discussed below). \ We will see
that in general the four solutions comprise two pairs, \ of which only one
pair is relevant to the system's behavior near the upstream boundary. \ The
two solutions of a pair together give one physical oscillation mode with an
arbitrary phase. \ 

\subsection{Pure FDO: $\delta_{V}=\delta_{D}=1$}

\begin{figure}
[ptb]
\begin{center}
\includegraphics[
height=3.2379in,
width=4.0145in
]%
{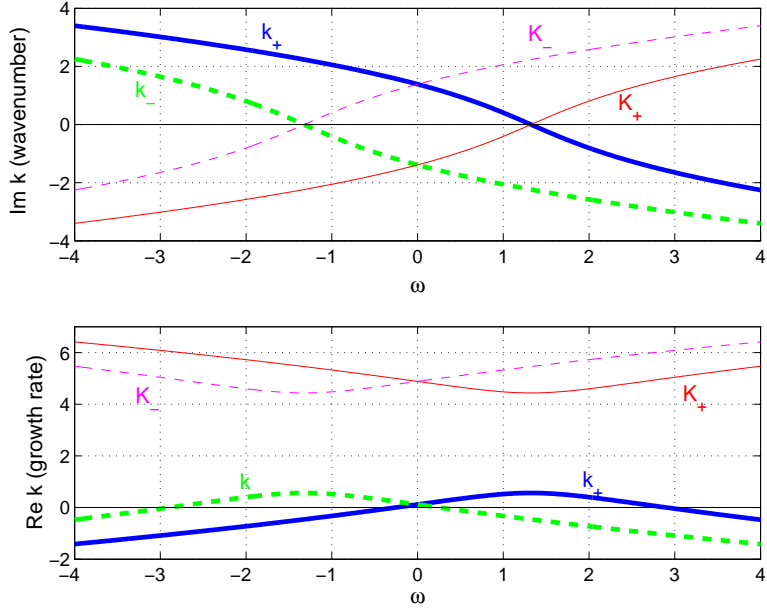}%
\caption{(color online) All four solutions of the dispersion relation for the
pure FDO case with a Hopf unstable local system and no differential transport.
\ The two lower solutions are labelled $k_{\pm}$ and the two upper solutions
are $K_{\pm}$. \ $K_{+}$ and $k_{+}$ are both associated with the eigenvector
$\mathbf{u}_{+}$ and the other two with $\mathbf{u}_{-}$. \ }%
\label{purefdo}%
\end{center}
\end{figure}

To illustrate the physical meaning of the dispersion relation, we consider
first the simplest case of pure FDO, in which $\delta_{V}=\delta_{D}=1.$ \ It
can easily be verified that in this case the dispersion relation
(\ref{dispersion}) relation does not depend on the Jacobian matrix elements
separately, but only on the trace and determinant. \ The pair of equations
(\ref{doubledisp}) can then be diagonalized completely by changing coordinates
to the eigenbasis of the Jacobian, and the quartic dispersion relation
factorizes into two quadratic ones as derived in \cite{McGraw}, one for each
eigenvector. \ The quadratic dispersion relations depend on the Jacobian
eigenvalues, which are given by $Tr/2\pm\sqrt{Tr^{2}/4-\Delta}$\ and are
complex conjugates\ if $Tr^{2}/4-\Delta<0$. \ \ For the sake of simplicity we
consider the particular case with
\begin{equation}
\mathbf{J}=%
\begin{pmatrix}
1 & 1\\
-1 & 1
\end{pmatrix}
,
\end{equation}
whose eigenvalues and eigenvectors are $\lambda_{\pm}=1\pm i$ and
\begin{equation}
\mathbf{u}_{\pm}\mathbf{=}%
\begin{pmatrix}
1\\
\pm i
\end{pmatrix}
\
\end{equation}
\ In general, the relative amplitude of the two components is complex for an
oscillatory system, indicating a phase difference between the the two
components. \ In this particular case, the phase difference is $\pi/2$. \ A
plot of the four solutions $k_{m}(\omega)$ ($m=1,2,3,4$) then has the form
shown in figure \ref{purefdo}. \ \ There are two solutions associated with
each of the two eigenvectors, \ of which one has a much larger real component.
\ \ Both solutions are necessary in order to satisfy a boundary value problem
in which boundary conditions are specified at $x=0$ and at some downstream
point $x=L$. \ However, \ it has been argued\cite{Proctor} that for quite
general boundary conditions at a far-away downstream boundary, \ it is the
solution with the \emph{less} positive growth rate that predominates near the
upstream boundary ($x\ll L$). \ \ As an example, \ consider adjusting the
coefficients $A$ and $B$ in the general solution $Ae^{\kappa_{1}x}%
+Be^{\kappa_{2}x}$ (with $\kappa_{2}>\kappa_{1}$) so as to satisfy either
Dirichlet or Neumann boundary conditions at $x=L$. \ \ In either of these
cases, $B$ is smaller than $A$ by a factor of order $e^{(\kappa_{2}-\kappa
_{1})L}$. \ Therefore, when there is a clear separation between the pairs of
solutions, the one with lower growth rate dominates everywhere but close to
the downstream boundary, which we take to be far away compared to the growth
or damping length scales$\ $($\left\vert \kappa_{1,2}\right\vert L\gg1$ ).
\ We therefore focus attention on the lower solutions with smaller growth
rates. \ These two solutions are associated with the two eigenvectors of the
Jacobian, so we may label them $k_{\pm}$. \ \ They are complex conjugate
mirror images of each other under reflection through the vertical axis
($k_{+}(\omega)=k_{-}^{\ast}(-\omega)$) \ and represent the same physical wave
solution, namely%
\begin{equation}%
\begin{pmatrix}
a\\
b
\end{pmatrix}
=\operatorname{Re}\left[  e^{i\omega t}e^{k_{+}x}\mathbf{u}_{+}\right]
=\operatorname{Re}\left[  e^{-i\omega t}e^{k_{-}x}\mathbf{u}_{-}\right]  =%
\begin{pmatrix}
\cos(\omega t+\operatorname{Im}k_{+}x)\\
\sin(\omega t+\operatorname{Im}k_{+}x)
\end{pmatrix}
e^{(\operatorname{Re}k_{+})x}.\label{physicalsolution}%
\end{equation}
Because of this reflection symmetery, it will be convenient in the remainder
of the paper to plot only one solution, with the understanding that the
reflected complex conjugate is also present. \ 

Figure \ref{purefdo} shows the growth rates and wave numbers for all four
solutions. \ \ Note that: \ 1) $\operatorname{Im}k$ has a zero at the natural
frequency $\omega_{0}=\beta$. \ \ When $\operatorname{Im}k$=0, the phase
velocity%
\begin{equation}
c=-\omega/\operatorname{Im}k\label{phasevelocitydef}%
\end{equation}
has a corresponding pole, as will be seen in figures \ref{difi1}%
-\ref{difituring}. \ Disturbances at precisely the natural frequency result in
growing uniform oscillations of the medium (this is essentially the batch Hopf
mode) while perturbations faster or slower than $\beta$ give downstream or
upstream travelling waves, respectively \cite{Kaern1}\cite{Faraday}.
\ Stationary waves occur for $\omega=0$. \ \ \ As shown in \cite{McGraw}, the
sharpness of the growth rate peak depends on the dimensionless quantity
$DTr/\phi^{2}$. \ Increasing the value of $D$ \ (or, equivalently, reducing
the flow velocity) \ makes the peak sharper and narrower and reduces the gap
between the upper and lower solutions, until, at the threshold of absolute
instability, the growth rate curve develops a cusp and the upper and lower
solutions cross. \ When the growth rate curves of the two solutions cross, it
is no longer correct to view the solution in the bulk as being determined
primarily by the upstream boundary condition--- it is also strongly influenced
by downstream conditions. \ This is the signature of an absolute rather than a
convective instability. \ \ The peak of the growth rate occurs precisely at
the frequency defined by the imaginary part of the Jacobian eigenvalue. \ If
the fixed point is not Hopf unstable but instead has eigenvalues $\alpha\pm
i\beta$ with $\alpha=Tr/2\leq0$, then the picture is qualitatively the same,
except that the peak remains below the horizontal axis. \ \ Thus all
perturbations are damped in this case, \ but the most slowly damped ones are
at the natural frequency. \ \ \ 

The pure FDO case can be viewed as a \textquotedblleft
baseline\textquotedblright\ for the physical interpretation of the dispersion
relations and their solution curves in the presence of differential transport.
\ Differential transport will modify the shapes of the curves, and will make
the eigenvectors $\omega$-dependent and no longer coincident with those of the
Jacobian. \ 

Finally, note that in the case when the fixed point is an unstable node rather
than a focus, the Jacobian has distinct real eigenvalues and eigenvectors
instead of complex conjugate pairs. Peak growth rates for the modes along both
eigenvectors then occur at $\omega=0.$\cite{McGraw} \ The above results are
qualitatively universal for any system with a Hopf instability. \ %

\subsection{The effects of differential transport}

\begin{figure}
[ptb]
\begin{center}
\includegraphics[
height=3.4973in,
width=4.1485in
]%
{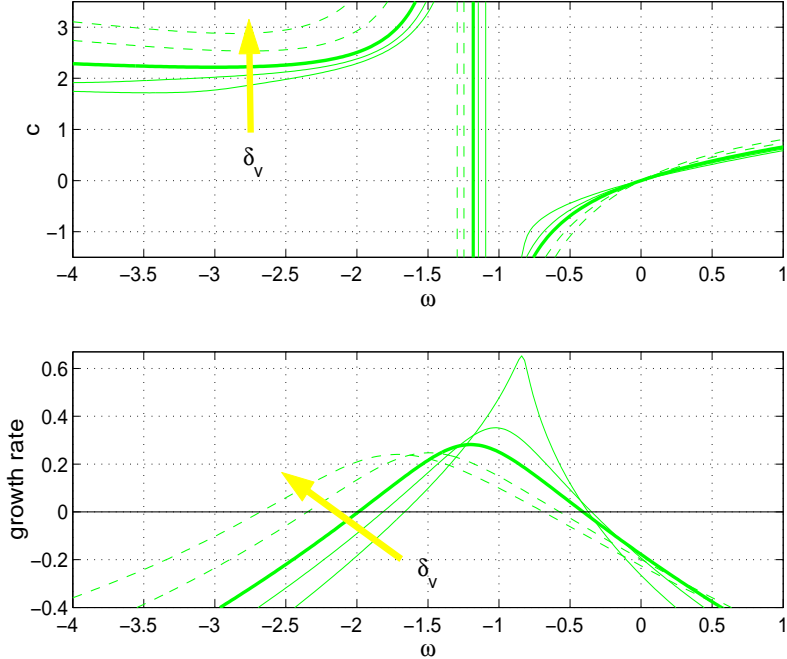}%
\caption{(color online) Effect of differential flow on the FDO peak. \ FN
model, $\varepsilon=1.5,D=0.1,\delta_{D}=1.$ \ $\delta_{v}=0.5,0.75,1,1.5,2.$
For the family of curves, the arrows show the direction of increasing
$\delta_{v}$. \ Dashed lines: \ $\delta_{v}>1$; solid thin lines: $\delta
_{v}<1$; \ Thick line: \ $\delta_{v}=1$.}%
\label{difi1}%
\end{center}
\end{figure}

\begin{figure}
[ptbptb]
\begin{center}
\includegraphics[
height=3.5898in,
width=4.4002in
]%
{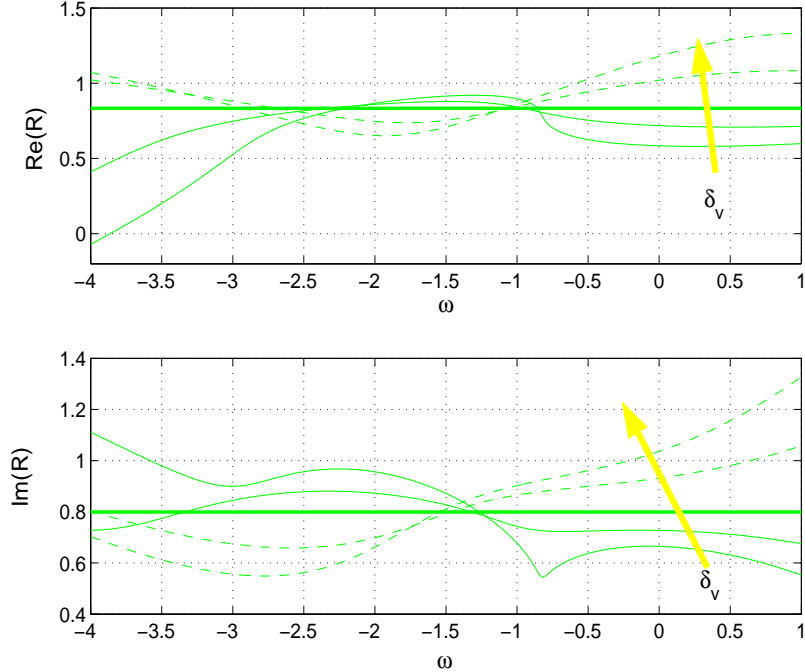}%
\caption{(color online) Amplitude ratios $R$ plotted for the same cases as in
figure \ref{difi1}. \ For the pure FDO case ($\delta_{v}=1$; thick line) the
amplitude ratio is constant, corresponding to an eigenvector of the Jacobian.
\ In the presence of differential transport, however, it becomes
frequncy-dependent. \ The imaginary part of the amplitude ratio is related to
the relative phase of the two components. \ }%
\label{Difi2}%
\end{center}
\end{figure}

\begin{figure}
[ptbptbptb]
\begin{center}
\includegraphics[
height=3.6071in,
width=4.235in
]%
{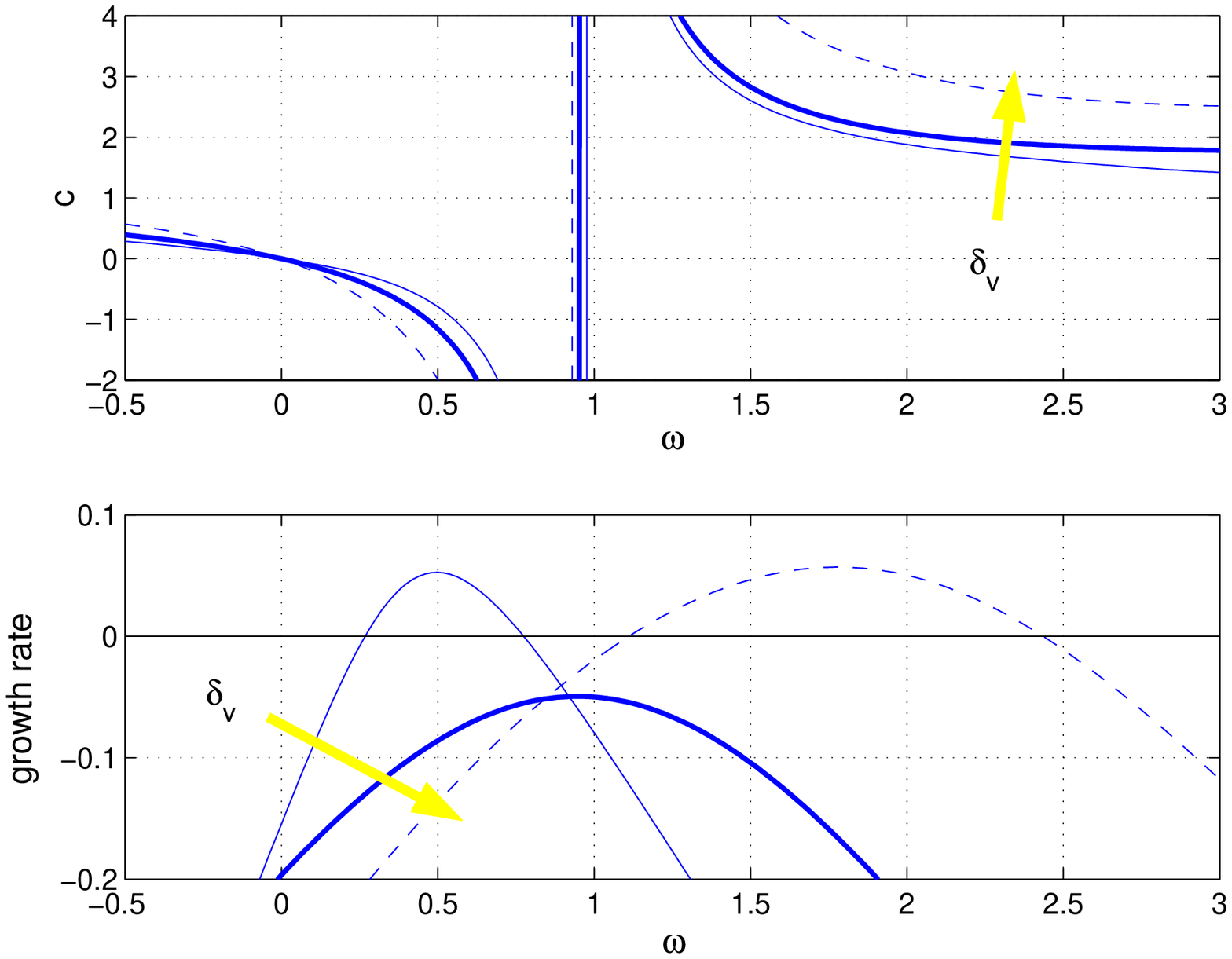}%
\caption{(color online) An example of the differential flow instability. \ FN
model, $\varepsilon=0.9$, $D=0.2$, $\delta_{D}=1$, $\delta_{v}=0.5$ (thin
solid line), 1 (thick line), 2 (dashed line). \ In the absence of differential
flow, \ there are no unstable modes, but there is nonetheless a peak in the
(negative) growth rate at the natural oscillation frequency. \ Sufficiently
strong differential flow shifts the peak so that it rises above zero and
unstable travelling wave modes appear. \ For $\delta_{v}>1$, the unstable
modes are downstream travelling waves, for $\delta_{v}<1$ they are upstream.
\ The medium remains stable against uniform oscillations. \ Viewed in this
way, the differential flow instability can be viewed as simply a continuous
deformation of the FDO instability. \ }%
\label{puredifi}%
\end{center}
\end{figure}

\begin{figure}
[ptb]
\begin{center}
\includegraphics[
height=3.6149in,
width=4.1995in
]%
{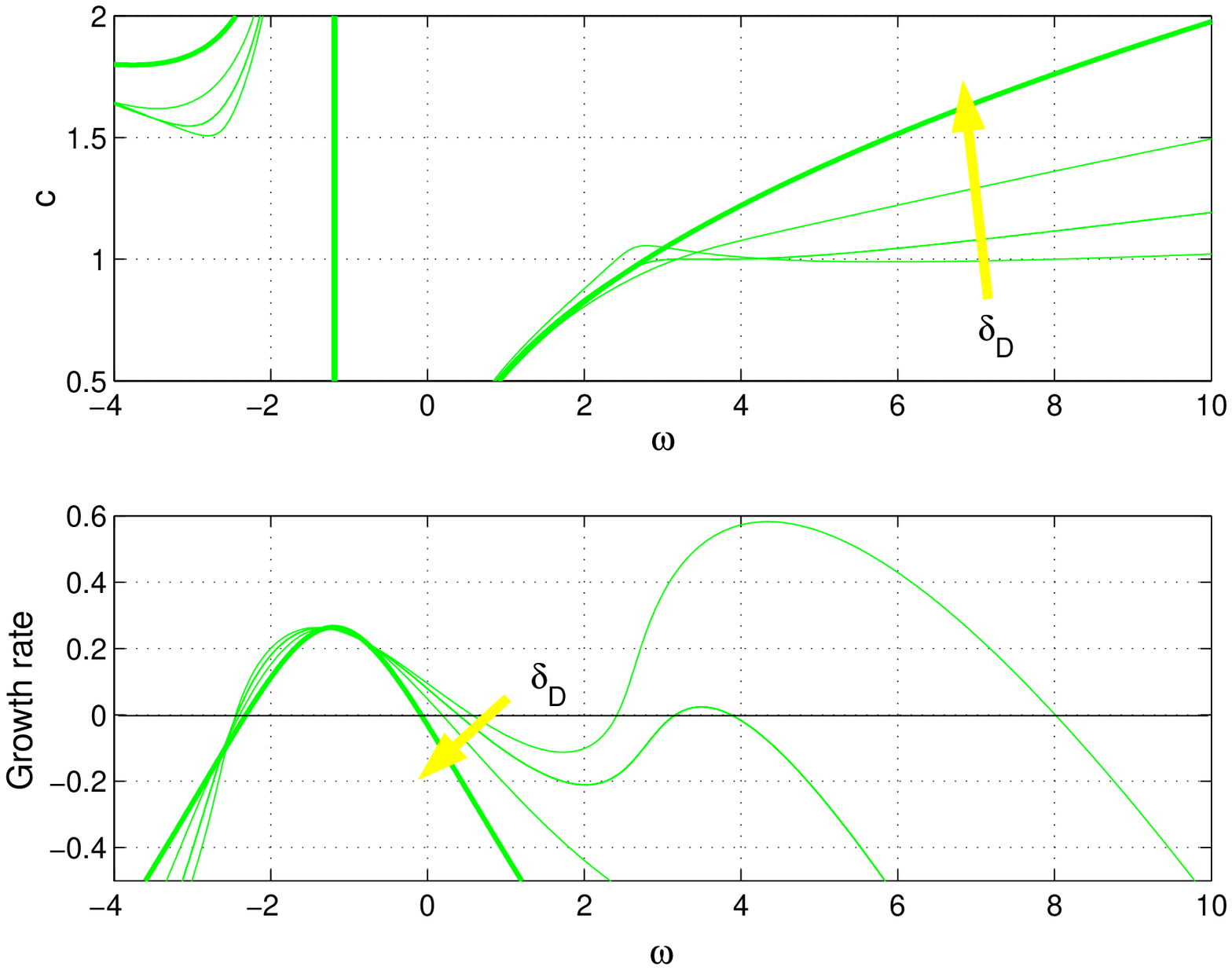}%
\caption{(color online) The effect of differential diffusion on the dispersion
relation in the absence of differential flow. \ FN model, $\varepsilon
=1.5,D=0.2,\delta_{v}=1$. \ $\delta_{D}=0.1,0.25,0.5,1.$ \ Thick line:
$\delta_{D}=1$; thin lines: $\delta_{D}<1$. \ Arrows show the direction of
increasing $\delta_{D}$. \ As $\delta_{D}$ decreases, the FDO peak is
distorted slightly, but much more noticeable is the growth of a second peak.
\ The modes in this peak can be identified as Turing modes. \ Their phase
velocity is close to 1, and, as shown in fig \ref{turing2}, their amplitude
ratio is almost purely real. \ }%
\label{turing1}%
\end{center}
\end{figure}

\begin{figure}
[ptbptb]
\begin{center}
\includegraphics[
height=3.5604in,
width=4.1831in
]%
{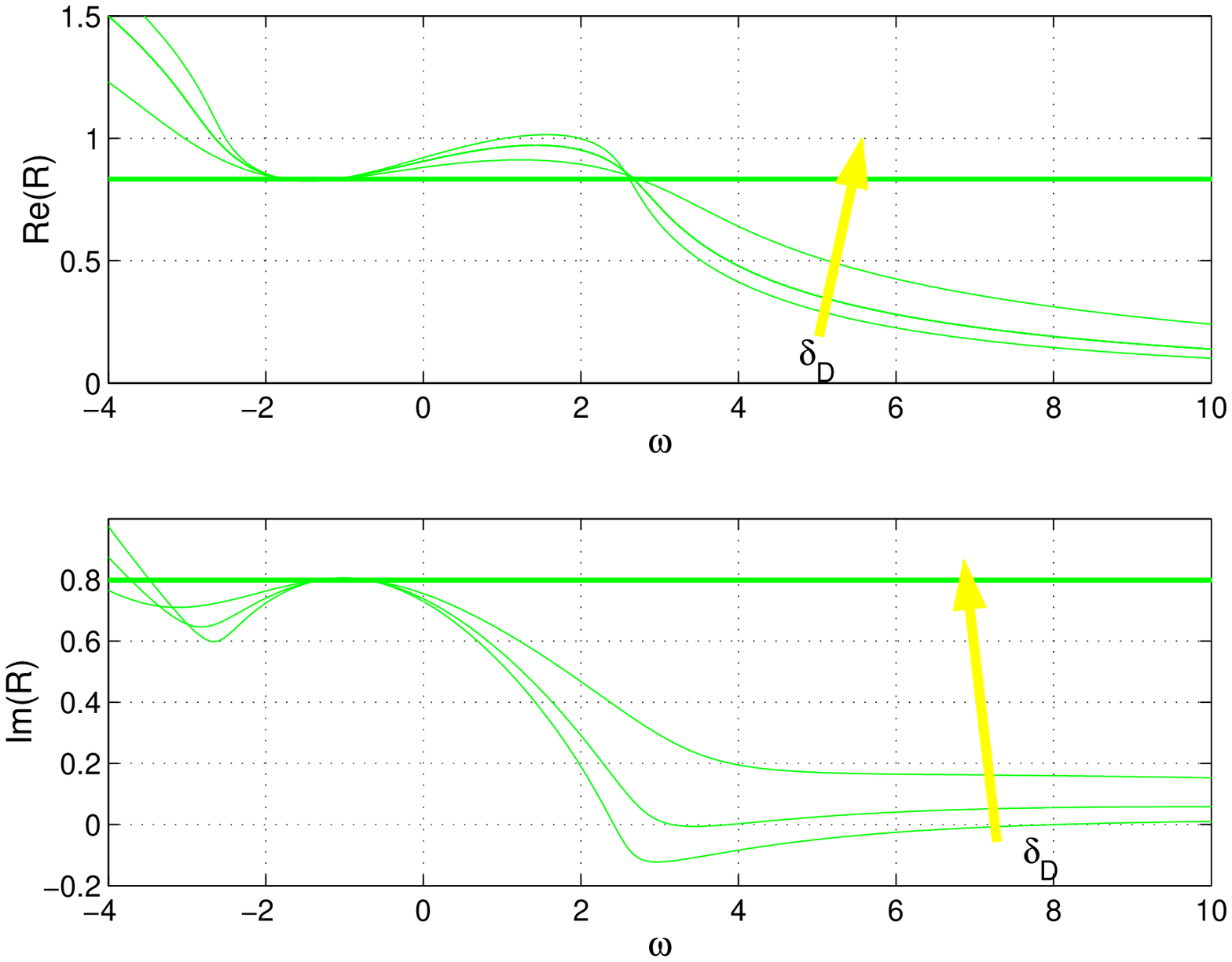}%
\caption{(color online) Amplitude ratio $R$ plotted for the same cases as in
figure \ref{turing1}. \ The imaginary component nearly vanishes for
frequencies within the Turing peak. \ Near the FDO peak, on the other hand,
the amplitude ratio is close to that of an eigenvector of the Jacobian. \ }%
\label{turing2}%
\end{center}
\end{figure}

\begin{figure}
[ptb]
\begin{center}
\includegraphics[
height=3.5604in,
width=4.1485in
]%
{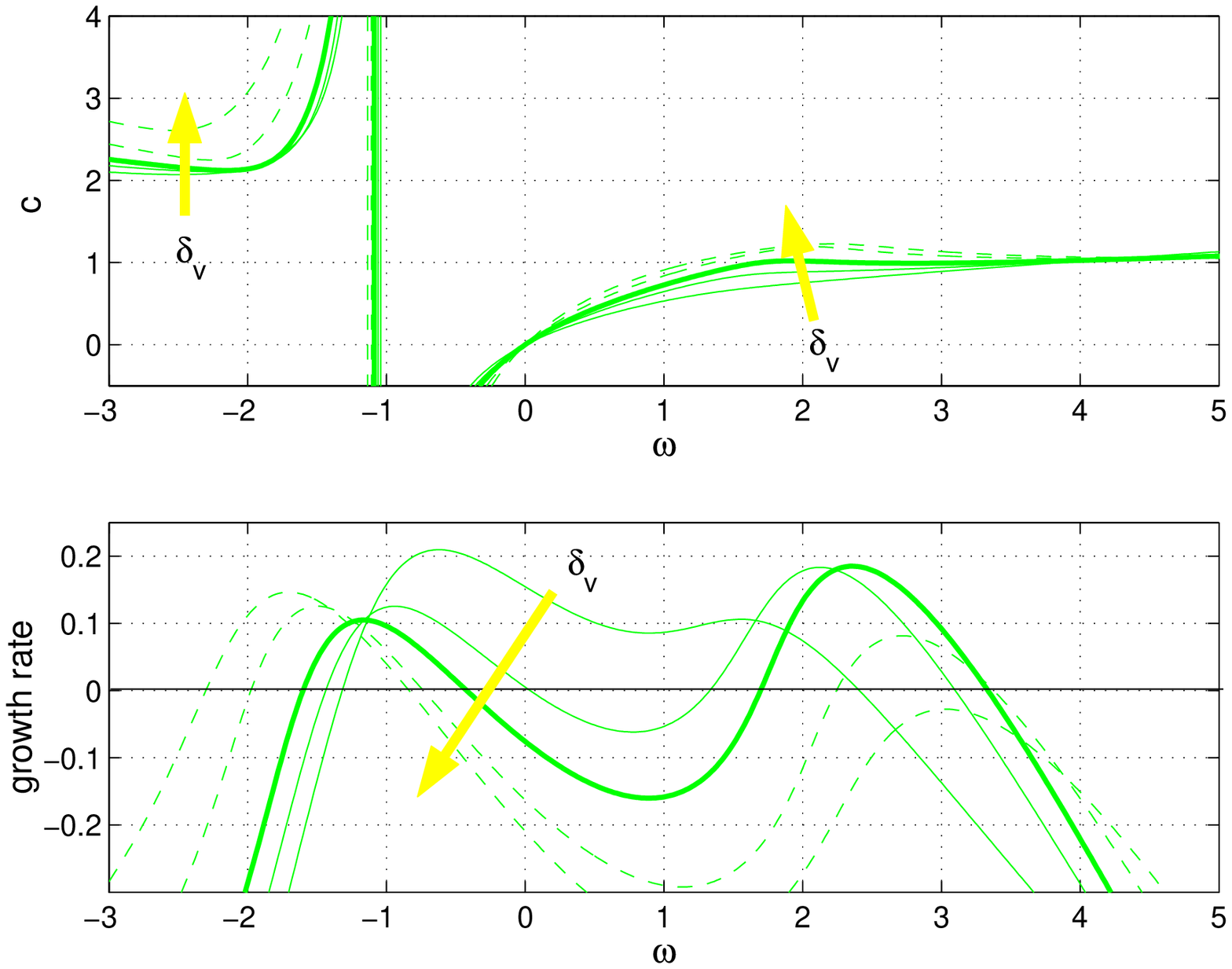}%
\caption{(color online) Interacting effects of differential flow and
diffusion. \ Here the differential diffusion is constant, and the velocity
ratio is varied. \ FN model, $\varepsilon=1.2,D=0.05,\delta_{D}=0.15.$
$\delta_{v}=0.5,0.75,1,1.5,2.$ \ Thick curve: $\delta_{v}=1$; thin solid
curves: $\delta_{v}<1$; dashed curves: $\delta_{v}>1$. \ With increasing flow
ratio, the two peaks are shifted closer together and begin to merge.
\ Departure from equal flow in either direction raises the height of the FDO
peak while lowering the Turing peak. \ Departure from equal flow also causes
the phase velocities of the Turing modes to depart from 1. \ }%
\label{difituring}%
\end{center}
\end{figure}

\begin{figure}
[ptbptb]
\begin{center}
\includegraphics[
height=3.5604in,
width=4.0906in
]%
{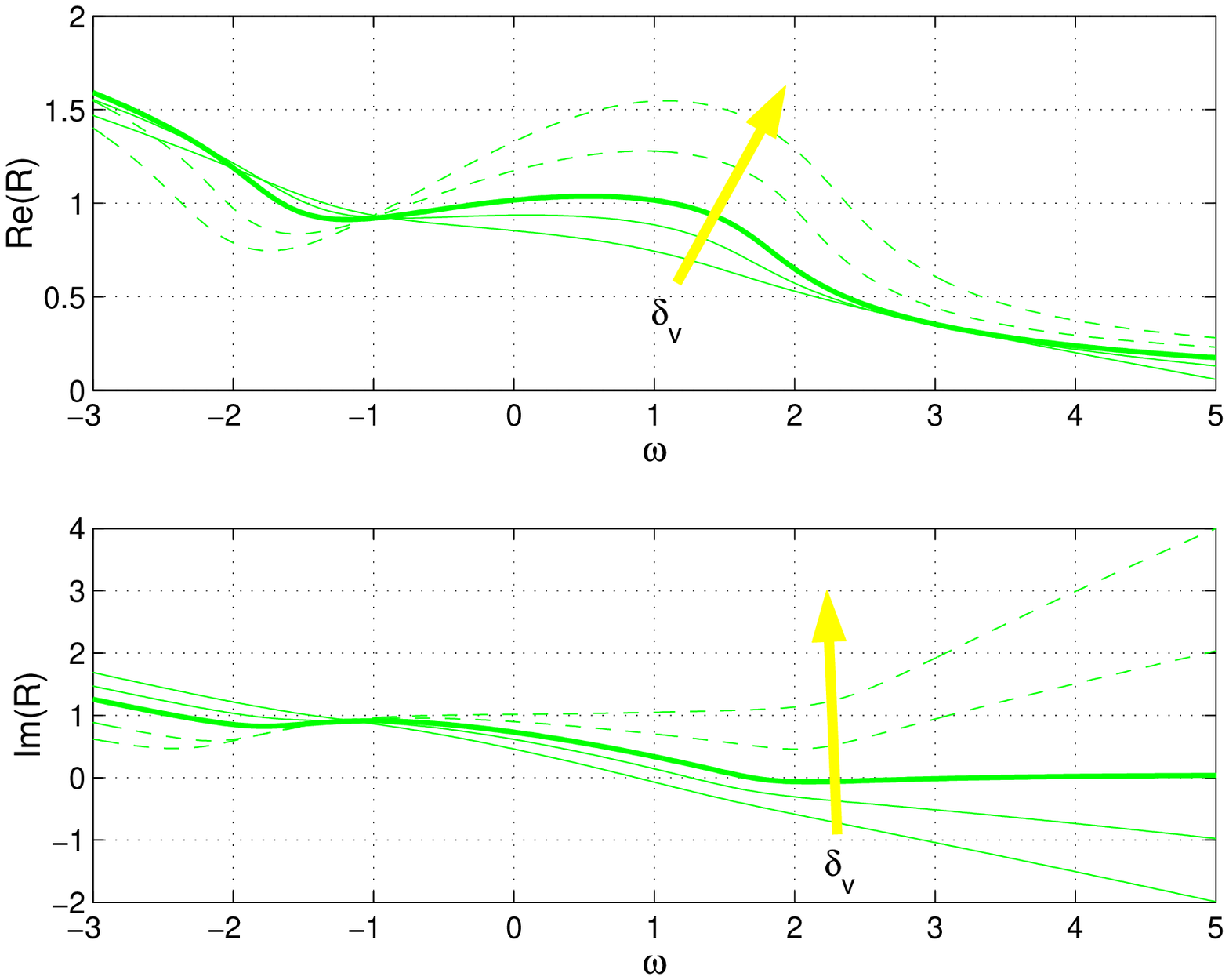}%
\caption{(color online) Amplitude ratios plotted for the same cases as in fig.
\ref{difituring}}%
\label{difituring2}%
\end{center}
\end{figure}

With the pure FDO case as a comparison, we now examine the effects of
differential transport and the convective, boundary driven manifestations of
DIFI and Turing instabilities. \ \ \ Typical results are shown in figs.
\ref{difi1}-\ref{difituring2} using the Fitzhugh-Nagumo-like \cite{FNModel}
(FN) model (\ref{FNeqns}-\ref{FNJac}) for the local dynamics.

The key features of the relevant solutions in the FDO case are that $k_{+}$
has a growth peak and the associated phase velocity has a pole at positive
$\omega$, while $k_{-}$ has a peak and pole on the opposite side, $\omega<0$.
\ \ A brief summary of the effects of differential transport on the dispersion
relation solutions is as follows: \ \ 

1. The primary effect of differential flow is to displace and distort the
positive-$\omega$ peak of $k_{+\text{ }}$(and its mirror image in $k_{-}$).
\ Depending on the details of the model, the peak may be shifted to the left,
right, upward or downward. \ The pole in the phase velocity may also be
shifted left or right.\ \ If sufficiently strong, \ differential flow can
raise the peak growth rate from negative to positive, thus giving an
instability even for a stable fixed point. \ This is precisely what happens in
DIFI. \ For $\delta_{V}\neq1$, \ the peak growth rate may occur quite far from
the pole of the phase velocity; thus the strongest instability is to a
travelling wave solution rather than to a uniform oscillation. \ 

2. \ The most important effect of differential diffusion in the absence of
differential flow ($\delta_{D}\neq1,\delta_{v}=1$) is to alter the shape of
the negative-$\omega$ tail of the $k_{+}$ solution (or, equivalently, the
positive tail of $k_{-}$). \ For fast inhibitor diffusion ($\delta_{D}<1$) the
negative tail can develop first an inflection point and then a second growth
rate peak. \ \ The modes within this second peak are distinguished by the
following features, confirming their interpretation as Turing patterns imposed
by the boundary condition and advected with the flow: \ \ A) \ Their phase
velocities are all very close to unity (in units where the flow velocity is 1)
showing that they are stationary in the comoving frame. \ B) \ The amplitude
ratio $R$ from the associated eigenvector is almost purely real, meaning that,
contrary to the situation in self-sustained oscillations, \ there is no phase
lag between the two species (the activator and inhibitor are almost exactly in
phase or $\pi$ out of phase). \ 

For the most general case of differential transport, then, the dispersion
relation has either one or two peaks, which we can identify as Hopf/FDO/DIFI
and Turing peaks respectively. \ \ Changing $\delta_{D}$ can alter the shape
of the FDO peak and conversely, $\delta_{V}$ can alter the Turing peak, but
they generally retain a separate identity, and for the most part exist in a
\textquotedblleft see-saw\textquotedblright\ relation. \ 

We now illustrate these statements using examples based on particular models
for the form of the Jacobian. \ The examples we present in figures
\ref{difi1}-\ref{difituring2} and the nonlinear simulations used a form of the
FitzHugh-Nagumo (FN) model\cite{FNModel}, \ for which the local dynamics is
given by
\begin{align}
\frac{dA}{dt} &  =\varepsilon(A-A^{3}-B)\label{FNeqns}\\
\frac{dB}{dt} &  =-B+2A\nonumber
\end{align}
and the Jacobian is
\begin{equation}
\mathbf{J=}%
\begin{pmatrix}
\varepsilon & -\varepsilon\\
2 & -1
\end{pmatrix}
\label{FNJac}%
\end{equation}
where $\varepsilon$ is a control parameter. \ $A$ and $B$ play the roles of
activator and inhibitor, respectively. \ A Hopf bifurcation occurs at
$\varepsilon=1$; \ the fixed point is unstable for $\varepsilon>1$. \ \ We
also studied another model using the simpler Jacobian%
\begin{equation}
\mathbf{J}=%
\begin{pmatrix}
\alpha & 1\\
-1 & \alpha
\end{pmatrix}
\label{alphaJac}%
\end{equation}
with control parameter $\alpha$ and Hopf instability for $\alpha>0$,
\ hereafter referred to as the \textquotedblleft$\alpha$%
-model.\textquotedblright\ \ For the $\alpha$-model with $\alpha>0$, both
species are autocatalytic, but one inhibits the other.\ \ In most cases,
qualitatively similar results were obtained for both models. When the results
for the $\alpha$-model differ from theose of the FN model, we describe them verbally.

Figures \ref{difi1} and \ref{Difi2} show typical effects of differential flow
on the FDO peak. \ Differential flow with either fast inhibitor or fast
activator transport shifts the position and height of the peak growth rate.
\ It can also shift slightly the location of the pole in the phase velocity
(i.e., the zero of $\operatorname{Im}k$) \ but usually this shift is less
pronounced than the shift of the peak in $\operatorname{Re}k$. \ In the case
shown in figures \ref{difi1} and \ref{Difi2}, \ the peak height and location
are both apparently monotonic in $\delta_{v}$ near $\delta_{v}=1$; \ the peak
shifts upward and to the right as $\delta_{v}$ decreases. \ \ This is not
universal, however. \ In some cases (see for example figures \ref{puredifi}
and \ref{difituring} below), \ the peak height has its minimum when
$\delta_{v}=1$, so that faster flow of \emph{either} species raises the height
of the instability peak. \ In one case we examined using the marginally stable
$\alpha$-model with $\alpha=1$, the peak shifts to the right both for
$\delta_{v}>1$ and $\delta_{v}<1$. \ \ What has been referred to as the
differential flow instability (DIFI) \ can be understood a special case in
which a growth rate peak whose maximum is less than zero in the absence of
differential flow is shifted above zero when $\delta_{v}\neq1$, thus creating
a convective instability even though the fixed point of the local system is
stable. \ An example of this is shown in figure \ref{puredifi}. \ Physically,
the shifting of the peak relative to the phase velocity pole means that in the
presence of differential flow the fastest-growing mode is a travelling wave
with some finite velocity, rather than a uniform oscillation. \ From figure
\ref{Difi2} it is evident that, while the amplitude ratio $R$ of the two
species remains constant in the pure FDO case $\delta_{v}=1$, \ differential
flow modifies both their amplitude and phase relations in a
frequency-dependent manner. \ \ \ \ \ \ \

Figures \ref{turing1} and \ref{turing2} show the effect of differential
diffusion without differential flow $(\delta_{D}\neq1,\delta_{v}=1)$. \ \ A
family of solution curves is shown for $\delta_{D}\leq1$ (i.e., equal
diffusion or fast inhibitor diffusion). \ This is the case that renders a
Turing instability possible in a stationary medium. \ The growth rate peak is
distorted somewhat relative to that of the pure FDO case. \ This effect was
more pronounced in some other examples we studied. \ In one case, fast
inhibitor diffusion lowered and broadened the FDO peak slightly while fast
activator diffusion raised and sharpened the peak significantly. \ In any
case, however, the distortion of the FDO peak is rather less salient than the
growth of a second peak at a different frequency. \ When this second peak
rises above zero, the modes contained within it have two important features:
\ First, their phase velocity is close to 1. \ The phase velocity curves in
fig. \ref{turing1} flatten out at $c\approx1$ for the range of amplified
frequencies in the second peak. \ This means that, in the co-moving frame, the
waves are stationary. \ \ Second, for frequencies within the range of the
second peak, \ the imaginary part of the amplitude ratio $\operatorname{Im}R$
is very small, \ indicating a lack of phase lag between the two components.
\ \ Both of these observations are consistent with the Turing instability
caused by differential diffusion. \ Because diffusion is directionally
symmetric, a mechanism driven by differential diffusion cannot cause a phase
lag between the two components. \ Turing patterns are reflection-symmetric,
and stationary in the co-moving frame. \ In view of these observations we
attribute the second peak to Turing-like modes and refer to it as a Turing
peak. \ In this example it is quite clearly separated from the FDO peak, \ the
latter being characterized by strongly frequency dependent phase
velocities\ and a non-zero, imaginary component $\operatorname{Im}R$ of the
amplitude ratio. \ There is a range of frequencies between the two peaks for
which there are only damped modes. \ \ In some cases, however, the two peaks
can grow broader and almost merge, so that as the driving frequency changes,
the resulting waves change continuously from an FDO-like to a Turing-like
character. \ Even in such cases, the Turing modes are distinguishable by means
of their near-unity phase velocities and nearly real amplitude ratios. \ \

In figures \ref{difituring} and \ref{difituring2}, we examine the interaction
between differential flow and differential diffusion, allowing both
differential transport modes to operate simultaneously, as in refs.
\cite{Satnoianu1}\cite{Satnoianu2}\cite{RazvanNew}. \ Here we plot the
dispersion solutions for a constant value of $\delta_{D}$ as $\delta_{v}$
varies. \ $\delta_{D}$ is such that a well-defined Turing peak exists for
$\delta_{v}=1$. \ \ We observe that setting $\delta_{v}\neq1$ shifts the FDO
peak as we expect. \ In this case, unlike that of figure \ref{difi1} but
similar to fig. \ref{puredifi}, the peak grows higher for either fast
activator or fast inhibitor flow. \ On the other hand, \ the Turing peak is
lowered for any $\delta_{v}\neq1$. \ The two peaks appear to have a
\textquotedblleft see-saw\textquotedblright\ relation. \ \ Differential flow
has other effects on the modes within the Turing peak. \ Their velocity begins
to depart from unity and is less uniform across the peak, and the amplitude
ratio is no longer purely real. \ In these senses, \ the \textquotedblleft
Turing\textquotedblright\ modes begin to lose their Turing-like character in
the presence of differential flow, even though one can still perceive two
separate peaks in the growth rate. \ \

\section{Nonlinear Simulations}

\begin{figure}
[ptb]
\begin{center}
\includegraphics[
height=2.7674in,
width=3.4852in
]%
{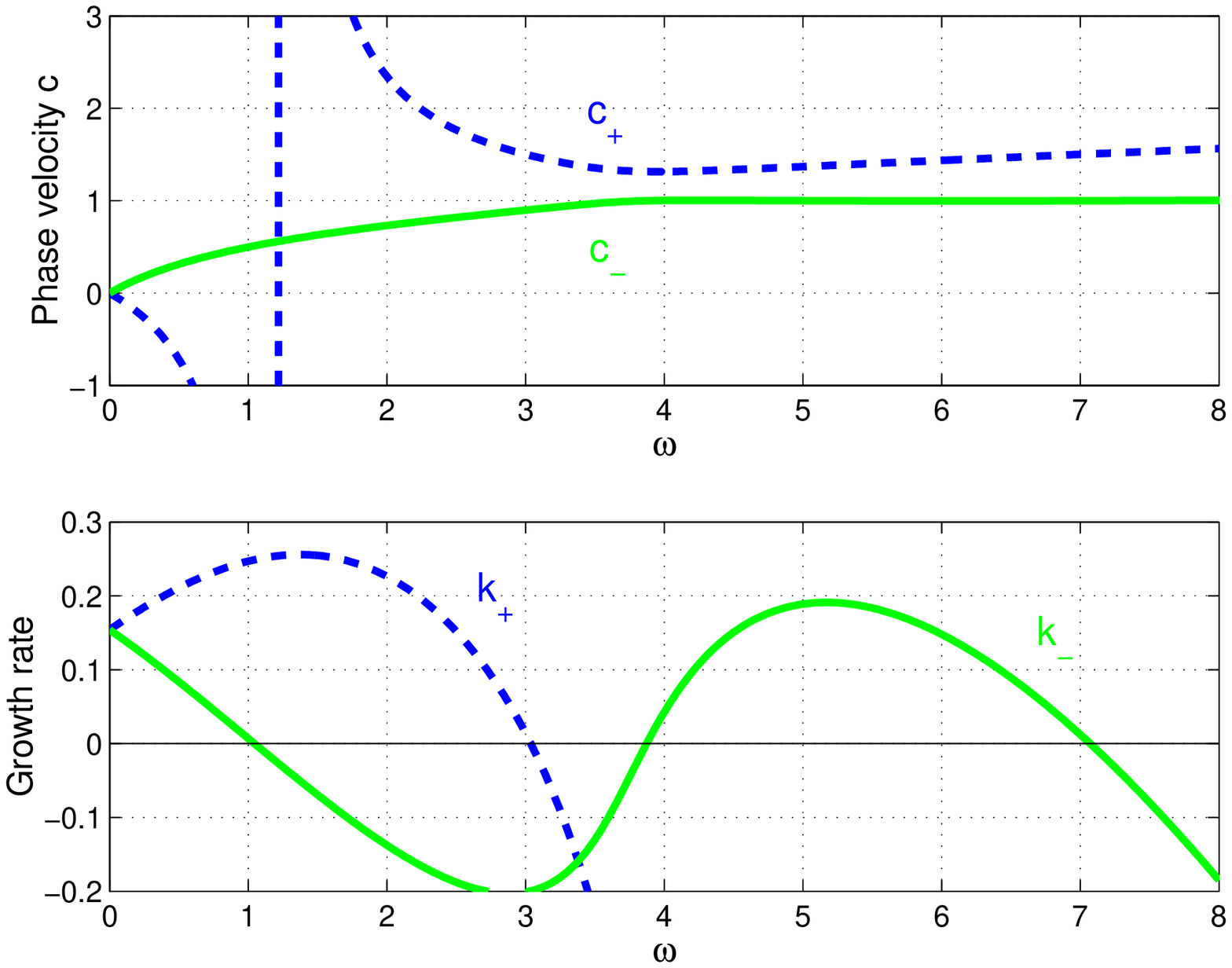}%
\caption{(color online) Growth rate and phase velocity as functions of
frequency for the model of our nonlinear simulations. \ Both relevant
solutions are plotted}%
\label{turingexdisp1}%
\end{center}
\end{figure}

\begin{figure}
[ptbptb]
\begin{center}
\includegraphics[
height=2.7674in,
width=3.493in
]%
{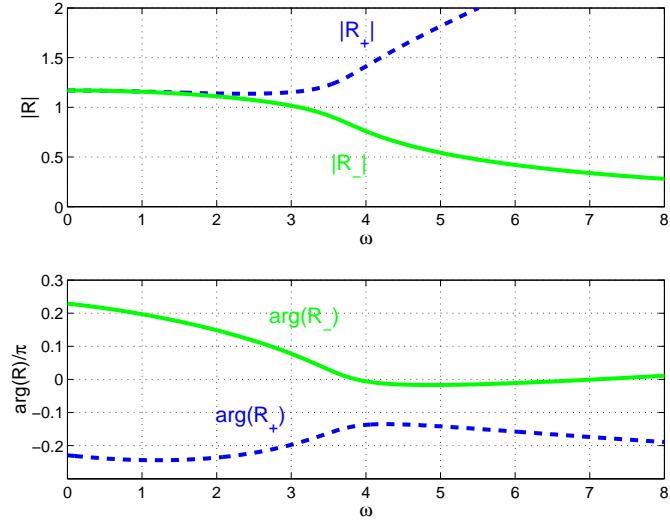}%
\caption{(color online) Modulus and phase of the complex amplitude ratio.
\ The modulus gives the ratio of the peak amplitudes for the oscillations of
the two dynamical variables, while the phase gives the relative phase shift.
\ }%
\label{turingexdisp3}%
\end{center}
\end{figure}

\begin{figure}
[ptbptbptb]
\begin{center}
\includegraphics[
height=7.5732in,
width=5.9802in
]%
{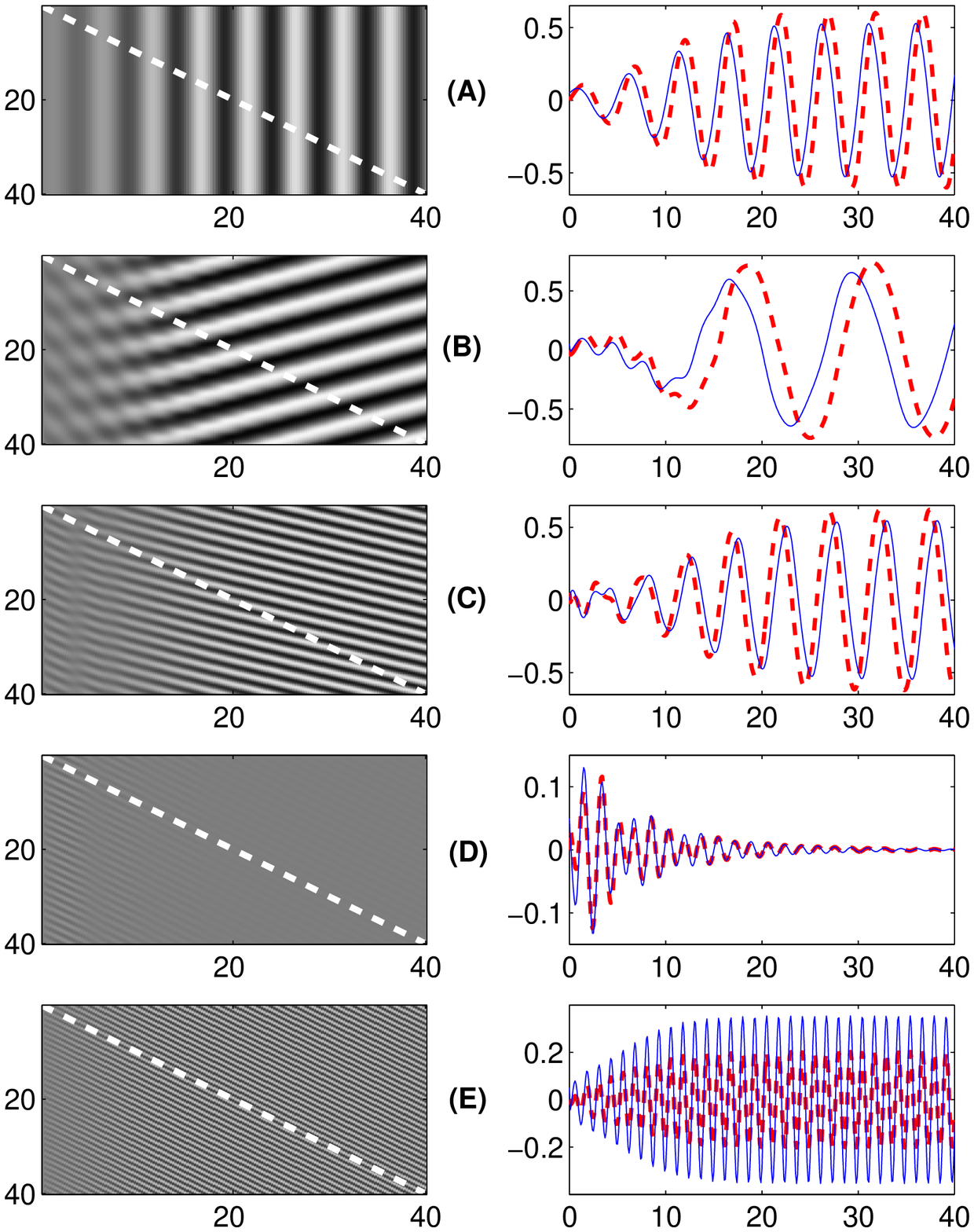}%
\caption{(color online): \ Results of nonlinear simulations of the FN model
for sinusoidal boundary perturbations at five different frequencies. \ a) 0 b)
0.9 c) 2.5 d) 3.5 \ e) 5. \ \ Left column: \ space(horizontal) vs. time
(vertical) plots with gray scale whowing concentration $A.$ Right column:
\ concentrations $A$ (thin line), $B$ (thick dashed line) vs. space
(horizontal axis). \ }%
\label{multiplots}%
\end{center}
\end{figure}

We now show the results of some nonlinear simulations of the FitzHugh-Nagumo
flow system in order to illustrate the application of the dispersion relations
to experiments. We choose to simulate the FN model with $\varepsilon=1.5$,
$D=0.05$, $\delta_{v}=1$ and $\delta_{D}=0.2$. \ As in the examples of figures
\ref{turing1} and \ref{turing2}, the dispersion relation shows both an FDO and
a Turing peak. \ The natural oscillation frequency (the pole in the phase
velocity) is $\omega_{0}\approx1.2$. \ \ For comparison with the simulations,
we plot both solutions $k_{\pm}$ on the same axes, for physical frequencies
$\omega>0.$ \ These plots are shown in figs. \ref{turingexdisp1}%
-\ref{turingexdisp3}. \ Figure \ref{multiplots} \ shows a series of
simulations with different driving frequencies. \ The boundary conditions for
these simulations were given by
\[
\mathbf{u}(0,t)=a_{0}%
\begin{pmatrix}
\cos\omega t\\
\sin\omega t
\end{pmatrix}
\]
where $a_{0}=0.05$. \ The plots in the left column of this figure show the
space-time patterns of the waves generated by the boundary perturbation. \ The
dotted white line in each plot represents the trajectory of a point co-moving
with the flow. This allows the phase velocities of the waves to be compared
readily with the flow velocity. \ The plots in the right column show both
dynamical variables $a$ and $b$ as functions of position for a single time.
The latter plots allow an examination of the waveforms, including the phase
shifts between activator and inhibitor. \ The dispersion relation (fig.
\ref{turingexdisp1}) predicts a positive growth rate at $\omega=0$, and,
accordingly, \ a constant (zero-frequency) perturbation indeed gives rise to
growing stationary waves which saturate at a finite amplitude. \ \ At
$\omega=0.9$, below the natural frequency, the dispersion relation shows that
both $k_{+}$ and $k_{-}$ have positive growth rates. $k_{-}$ (the solid curves
in fig. \ref{turingexdisp1}) gives waves with a phase velocity of $\sim0.5$,
i.e., \ downstream waves moving slower than the flow velocity, while $k_{+}$
gives waves with a negative phase velocity, i.e., upstream travelling waves.
\ \ Near the boundary, \ a superposition of both waves occurs, but the
upstream waves have a much larger growth rate and they dominate at positions
farther downstream, crowding out the other mode entirely and reaching a
nonlinear saturated amplitude. \ At $\omega=2.5>\omega_{0}$, \ only $k_{+}$
has a positive growth rate, giving waves with an positive (downstream) phase
velocity faster than the flow speed. \ A small admixture of the other mode
$k_{-}$ can be discerned near the boundary, but it decays rapidly with
downstream distance. \ \ $\omega=3.5$ falls within the gap between the FDO and
Turing peaks. \ Thus there are no growing modes at this frequency and the
disturbance decreases with downstream distance. \ \ $\omega=5$, however, lies
within the Turing peak of the $k_{-}$ solution. \ As predicted by the
dispersion relation, the resulting waves have a velocity nearly equal to the
flow velocity and there is almost no phase difference between the two species.
\

The amplitude and phase characteristics of the nonlinear waveforms may also be
compared with the predictions of the linear dispersion relation, and in this
case the agreement is quite close. \ In the complex exponential solution eq.
\ref{complexexp}, \ the modulus of the ratio $R=v/u$ gives the ratio of the
peak amplitudes of the oscillations of the two dynamical variables, while the
argument of $R$ gives the relative phase. \ For frequencies within the FDO
peak, \ our nonlinear simulations show that, as one expects from figure
\ref{turingexdisp3}, the ratio of the peak amplitudes of $b$ and $a$ is
slightly larger than unity, while the phase shift is approximately $\pi/4$.
\ For $\omega=5$, \ on the other hand, the linear dispersion relation gives a
very small phase shift for the relevant $k_{-}$ solution, and an amplitude
ratio slightly larger than $0.5$. \ The phase shift is indeed almost zero and
the $b$ amplitude is indeed smaller than that of $a$. \ The actual amplitude
ratio in the saturated waveform is approximately $0.6$, close to the
prediction of the linearized analysis.

\begin{figure}
[ptb]
\begin{center}
\includegraphics[
height=4.1321in,
width=5.271in
]%
{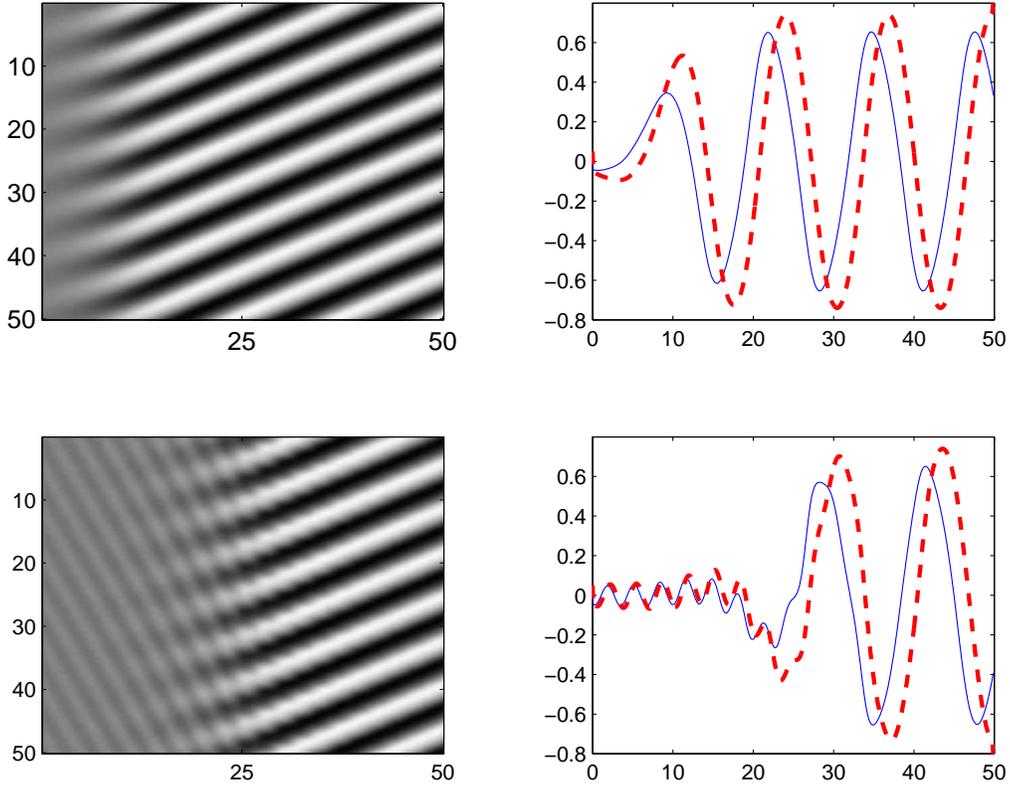}%
\caption{(color online) Simulations with sinusoidal boundary perturbations at
$\omega=0.9$, but with the driving amplitudes and phases tuned to select the
upstream (top) or downstream (bottom) mode. \ Even if the downstream mode is
selected near the boundary, the faster-growing upstream mode eventually takes
over farther downstream. \ As in fig. \ref{multiplots}, the plots on the left
are space-time, and those on the right are of $A$ and $B$ vs. position. \ }%
\label{selectionplot}%
\end{center}
\end{figure}

Finally, we note that at frequencies where more than one mode is present, one
can be selected by manipulating the driving function itself so as to align it
with one eigenvector or the other. \ As an example, consider $\omega=0.9$, a
frequency at which both solution branches exhibit positive growth rates.
\ Here, we find that for the faster-growing $k_{+}$ mode the complex amplitude
ratio is $R_{+}\approx0.84-0.8i$ while for the other mode $R_{-}%
\approx0.94+0.68i$. \ The driving function used in fig. \ref{multiplots}
excites both of these modes and a superposition is seen near the upstream
boundary. \ By tuning the driving function to be
\[
\mathbf{u}(0,t)=a_{0}\operatorname{Re}%
\begin{pmatrix}
1\\
R_{+}%
\end{pmatrix}
e^{i\omega t}=a_{0}\operatorname{Re}%
\begin{pmatrix}
e^{i\omega t}\\
\left\vert R_{+}\right\vert e^{i(\omega t+\phi_{+})}%
\end{pmatrix}
=a_{0}%
\begin{pmatrix}
\cos\omega t\\
0.84\cos\omega t+0.8\sin\omega t
\end{pmatrix}
,
\]
however, \ we can excite mostly the $k_{+}$ mode so that the upstream
travelling waves appear almost uncontaminated. \ Conversely, by choosing the
amplitude and phase of the driving to align with the other eigenvector, we
excite mostly the other, $k_{-}$ mode. \ Eventually, however, the other,
faster-growing mode begins to appear, possibly through nonlinear effects or
through the small admixture still present in the boundary condition, \ and the
$k_{+}$ mode eventually wins in the asymptotic downstream region. \ Simulation
results which show this selection effect are plotted in figure
\ref{selectionplot}. \

\section{Conclusions and experimental tests}

We have examined the linear stability analysis relevant to the convective
growth of spatiotemporal patterns in a reactive flow system, excited by a
small sinusoidal perturbation of a fixed point at the inflow boundary. We
examined the real and imaginary parts of the wavenumber as functions of the
real boundary forcing frequency. They represent the downstream growth rate and
periodicity of the disturbances caused by a boundary perturbation. \ We found
that the growth rate as a function of frequency exhibits at most two
physically distinct peaks corresponding to two types of waves, one or both of
which may be present and convectively unstable. We found that, in addition to
the phase velocity, the complex ratio $R$ of activator and inhibitor
concentrations, which encodes the relative amplitude and phase of oscillations
in the activator and inhibitor concentrations, provides an additional
criterion for distinguishing the types of modes. \ Turing-like modes are
distinguished from FDO-like modes by the lack of a phase lag between the two
components and by phase velocities that are close to the flow velocity, so
that in the co-moving frame they are stationary patterns. Viewing the
different types of modes as belonging to peaks in the growth rate, we saw the
close relationship between FDO and the differential flow instability. \ One
can be viewed as a continuous deformation of the other. \ A primary effect of
differential flow is to shift the FDO peak. \ What has been referred to as the
differential flow instability (the appearance of a travelling wave instability
in a medium which is neither Hopf nor Turing unstable) can be interpreted as a
case in which a sub-threshold FDO peak is shifted sufficiently to bring it
above zero and to create unstable modes (fig. \ref{puredifi}).

We now comment on experimental verifications of the present predictions. The
presence of two peaks could be seen in an experiment in which the perturbation
frequency at the inflow is the control parameter. \ Frequencies for which the
growth rate is positive will result in sustained waves, while the waves will
die out and fail to propagate if the growth rate is negative. \ Based on our
results we expect that growing waves will appear within at most two frequency
ranges. \ The phase velocities can also be measured and compared with our
general findings.

Two types of experiments may be envisaged, in which oscillatory driving is
implemented differently. \ The first type \cite{Kaern6}\cite{Miguez1}%
\cite{Santiago} makes use of a linearly growing, light-sensitive
reaction-diffusion system. \ The effective moving boundary is provided by a
moving mask which extinguishes the reaction on the illuminated side of a
moving line. \ \ The illumination at the moving boundary can be modulated
periodically, resulting in an oscillatory perturbation. In these experiments,
differential diffusion $\delta_{D}\neq1$ is achieved by immobilizing one
species on a gel, but differential flow is absent, $\delta_{v}=1$.

The second type of experiment is conducted in a packed bed reactor (PBR),
which is by the outlet of a continuous stirred tank reactor
(CSTR).\cite{Kaern1}\cite{Bamforth3}\cite{Toth}\cite{Kaern2}\cite{Kaern5}%
\cite{Rovinsky93} \ The CSTR may be manipulated to be stationary or to
oscillate slower or faster than the medium in the PBR. This leads to
stationary, upstream and downstream moving waves \cite{Kaern1}\cite{Kaern3}%
\cite{Kaern4}\cite{Kaern5}\cite{Faraday}. By packing the PBR with
ion-exchanger beads that immobilize either activator or inhibitor, conditions
of simultaneous differential diffusion and differential flow may be obtained.
\ Refs. \cite{Satnoianu1}\cite{Satnoianu2}\ modelled this by setting
$\delta_{D}=\delta_{v}.$ \ It is a challenge to devise experiments in which
the flow ratio, diffusion ratio and driving frequency can be varied
independently. While the phase velocities of travelling waves can easily be
measured, \ verification of other properties of the waves may present
experimental challenges. Verification of the predicted phase relationships
would require simultaneous measurements of both activator and inhibitor
concentrations. \

\end{document}